\documentclass[aps,prb,superscriptaddress,twocolumn]{revtex4}
\usepackage{graphicx}
\usepackage{color}
\newcommand{\nc}{\newcommand}
\nc{\black}{\color{black}}
\nc{\red}{\color{red}}
\nc{\blue}{\color{blue}}
\nc{\green}{\color{green}}
\nc{\yellow}{\color{yellow}}
\nc{\orange}{\color{orange}}
\nc{\violet}{\color{violet}}
\nc{\magenta}{\color{magenta}}
\nc{\grey}{\color{grey}}

\begin{document}
\title{The Origin of Magnetic Interactions in Ca$_3$Co$_2$O$_6$}
\author{Raymond Fr\'esard}
\email[]{Raymond.Fresard@ismra.fr}
\affiliation{Laboratoire Crismat, UMR CNRS-ENSICAEN(ISMRA) 6508, Caen, France}
\author{Christian Laschinger}
\affiliation{EPVI, Center for Electronic Correlations and Magnetism,  Augsburg University}
\affiliation{Laboratoire Crismat, UMR CNRS-ENSICAEN(ISMRA) 6508, Caen, France}
\author{Thilo Kopp}
\affiliation{EPVI, Center for Electronic Correlations and Magnetism,
Augsburg University}
\author{Volker Eyert}
\affiliation{Theoretical Physics II,
Augsburg University, 86135 Augsburg, Germany}
\date{\today}

\begin{abstract}
We investigate the microscopic origin of the ferromagnetic and
antiferromagnetic spin exchange couplings in the quasi one-dimensional
cobalt compound Ca$_3$Co$_2$O$_6$. In particular, we establish
a local model which stabilizes a ferromagnetic alignment of the $S=2$
spins on the cobalt sites with trigonal prismatic symmetry,
for a sufficiently strong Hund's rule coupling on the cobalt
ions. The exchange is mediated through a $S=0$ cobalt ion at
the octahedral sites of the chain structure. We present a strong
coupling evaluation of the Heisenberg coupling between the $S=2$ Co
spins on a separate chain. The chains are coupled antiferromagnetically 
through super-superexchange via short O-O bonds.

\end{abstract}

\maketitle
Recently there has been 
renewed interest in systems exhibiting
magnetization steps. In classical systems such as CsCoBr$_3$ 
one single plateau is
typically observed in the magnetization versus field curve at one
third of the magnetization at saturation.\cite{Hida94} This phenomenon
attracted considerable attention, and 
Oshikawa, Yamanaka and Affleck demonstrated that Heisenberg
antiferromagnetic chains
exhibit such magnetization plateaus when
embedded in a magnetic field.\cite{Oshikawa97} These steps are
expected when $N_c (S-m)$ is an integer, where $N_c$ is the number of
sites in the magnetic unit cell, $S$ the spin quantum number, and $m$ the
average magnetization per spin, which we shall refer to as the OYA
criterion. 
The steps can be stable when chains 
are coupled, for instance in a ladder geometry. In that case the
magnetic frustration is an important ingredient to their
stability.\cite{Mila98} 
Plateaus according to the OYA criterion are also anticipated for
general configurations, provided gapless excitations do not destabilize
them.\cite{Oshikawa00} Indeed several systems exhibiting magnetization
steps are now known;\cite{Shiramura98,Narumi98} they all obey the OYA
criterion, they are usually far from exhausting all the possible $m$ values, 
they all are frustrated systems, and they all can be described by an
antiferromagnetic Heisenberg model. Related behavior has been recently
found in other 
systems. For example, up to five plateaus in the magnetization
vs. field curve have been observed in Ca$_3$Co$_2$O$_6$ at low
temperature\cite{Aasland97,Kageyama97,Maignan00}.  
However there is to date no 
microscopic explanation to this phenomenon, even though the location 
of the plateaus is in agreement with the OYA criterion. 

\begin{figure}[h]
\centerline{\includegraphics[width=0.49\textwidth]{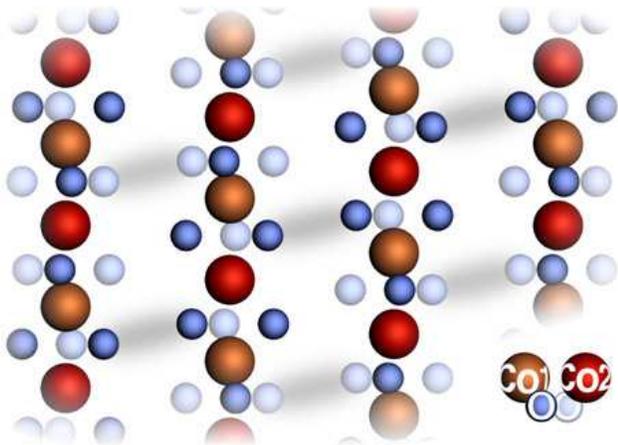}}
\caption{
Projection of Ca$_3$Co$_2$O$_6$ in the (100) plane, with
those oxygen atoms which are closest to the plane. 
Shadows indicate the shortest O-O bonds, along which super-superexchange
processes take place. Yellow (red) large spheres denote Co1 (Co2) atoms,
dark (light) small blue spheres oxygen atoms located above
(below) the Co plane.
\label{Fig:plane}}
\end{figure}

Ca$_3$Co$_2$O$_6$ belongs to the wide family of compounds
A'$_3$ABO$_6$, and its structure belongs to the space group R\=3c. 
It consists of infinite chains 
formed by alternating face sharing AO$_6$ trigonal prisms and
BO$_6$ octahedra --- where Co atoms occupy both A and B sites. Each
chain is surrounded by six chains separated by Ca atoms. As a result a
Co ion has two 
neighboring Co ions on the same chain, at a
distance of $2.59$~\AA, and twelve Co neighbors on the neighboring chains
at distances $7.53$~\AA\  (cf.\ 
Fig.~\ref{Fig:plane}).\cite{Fjellvag96} 
Concerning the magnetic structure, the experiment points toward a 
ferromagnetic ordering of the magnetic Co ions along the chains, together with 
antiferromagnetic correlations
in the buckling a-b plane.\cite{Aasland97} 
The transition into the ordered state is
reflected by a cusp-like singularity in the specific heat at
25~K,\cite{Hardy03} 
--- at the temperature where a strong increase of 
the magnetic susceptibility is observed.
Here we note that it is particularly
intriguing to find magnetization steps in a system where the
dominant interaction is ferromagnetic. 

In order to determine the effective magnetic Hamiltonian of a
particular compound one typically uses the
Kanamori-Goodenough-Anderson 
(KGA) rules\cite{Goodenough}. 
Knowledge of the ionic configuration
of each ion allows to estimate the various magnetic couplings. 
When applying this program to Ca$_3$Co$_2$O$_6$ one faces a
series of difficulties 
specifically when one tries to reconcile the 
neutron scattering measurements that each second Co ion is
non-magnetic. 
Even the assumption that every other Co ion is in a high spin state
does not settle the intricacies related to the magnetic properties;
one still has to challenge issues such as:
i) what are the ionization degrees of the Co ions?
ii) how is an electron transfered from one cobalt ion to a second? 
iii) which 
of the magnetic Co ions are magnetically coupled?
iv) which mechanism generates a ferromagnetic coupling along the chains?

These questions 
are only partially resolved by ab initio calculations. 
In particular, one obtains that 
both Co ions are in 3+ configurations.\cite{Whangbo03} 
Moreover both Co-O and direct Co-Co hybridizations
are unusually large, and 
low spin and high spin configurations for the Co ions along the chains
alternate.\cite{Eyert03}

Our publication addresses the magnetic couplings, and in particular
the microscopic origin of the ferromagnetic coupling of two Co ions
through a non-magnetic Co ion. In view of the 
plethoric variety of
iso-structural compounds,\cite{Stitzer01} the presented mechanism is
expected to apply to
many of these systems. We now derive the magnetic inter-Co coupling 
for Ca$_3$Co$_2$O$_6$ from microscopic considerations. 
The high-spin low-spin scenario confronts us with the question 
of how a ferromagnetic coupling can establish itself, taking into 
account that the high spin Co ions are separated by over 5~\AA, 
linked via a non-magnetic Co and several oxygens. 

Let us first focus on the Co-atoms in a single Co-O chain of
Ca$_3$Co$_2$O$_6$. As mentioned above the surrounding oxygens 
form two different environments in an alternating pattern. 
We denote the Co ion in the center of the oxygen-octahedron Co1, and 
the Co ion in the trigonal prisms Co2.
The variation in the oxygen-environment leads to three important
effects. First, there is a difference in the strength of the crystal
field splitting, being larger in the octahedral environment. As a
result Co1 is in the low spin state and Co2 in the high spin
state. Second, the local energy levels are in a different
sequence. For the octahedral environment we find the familiar 
$t_{2g}$--$e_g$
splitting, provided the 
axes of the local reference
frame point towards the surrounding oxygens. The trigonal prismatic
environment accounts for a different set of energy levels.
For this local symmetry one expects a level scheme with $d_{3z^2-r^2}$ as 
lowest level, followed by two twofold degenerate pairs 
$d_{xy}$, $d_{x^2-y^2}$ and $d_{xz}$, $d_{yz}$.
However, our LDA calculations\cite{Eyert03} show that the $d_{3z^2-r^2}$ 
level is actually slightly above the first pair of levels.
Having clarified the sequence of the energy levels, we now turn
to the microscopic processes which link the Co ions. Two
mechanisms may be competing: either the coupling 
involves the intermediate oxygens,
or direct Co-Co overlap is more important. Relying on electronic structure 
calculations, we may safely assume that 
the direct Co-Co overlap dominates.\cite{Eyert03} 
The identification of the contributing orbitals
is more involved. 
Following Slater and Koster\cite{Slater54} 
one finds that only the $3z^2$-$r^2$ orbitals along the chains have
significant overlap. However, we still have to relate
the Koster-Slater coefficients and the coefficients for the 
rotated frame  since the natural reference
frames for Co1 and Co2 differ. On the Co2 atoms with the triangular prismatic
environment the $z$-axis is clearly defined along the chain
direction, and we choose the $x$ direction to point toward one
oxygen. This defines a reference frame $S$. 
The $x$ and $y$ directions are arbitrary and irrelevant to our 
considerations. The octahedral environment surrounding the Co1 atoms
defines the natural coordinate system, which we call $S'$. 
By rotating $S'$ onto $S$ one obtains the $3z^2$-$r^2$ orbital in
the reference frame $S$ as an equally weighted sum of $x'y'$, $x'z'$,
$y'z'$ orbitals in $S'$. The above observation that 
the only significant overlap is due to the $3z^2$-$r^2$ orbitals
on both Co ions now translates into an overlap of the $3z^2$-$r^2$ orbital on high spin cobalt 
with all $t_{2g}$ orbitals on low spin cobalt.

\begin{figure}[t]
\centerline{\includegraphics[width=.47\textwidth,clip,angle=0]{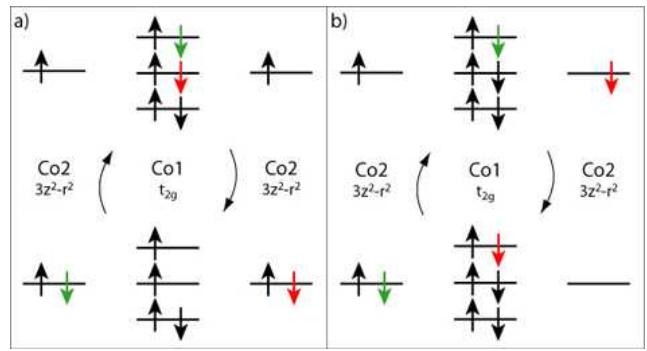}}
\caption{Typical hopping paths for a) ferromagnetic and b)
antiferromagnetic ordering. The displayed ferromagnetic path is
the only one for ferromagnetic ordering and has the highest
multiplicity of all, ferromagnetic and antiferromagnetic. 
There are similar paths for antiferromagnetic ordering but 
with a Hund's rule penalty and lower multiplicity. The path 
in (b) is unique for the antiferromagnetic case and has low energy 
but also low multiplicity. \label{Fig:levelscheme}}
\end{figure}

\begin{figure}[t]
\centerline{\includegraphics[width=0.48\textwidth]{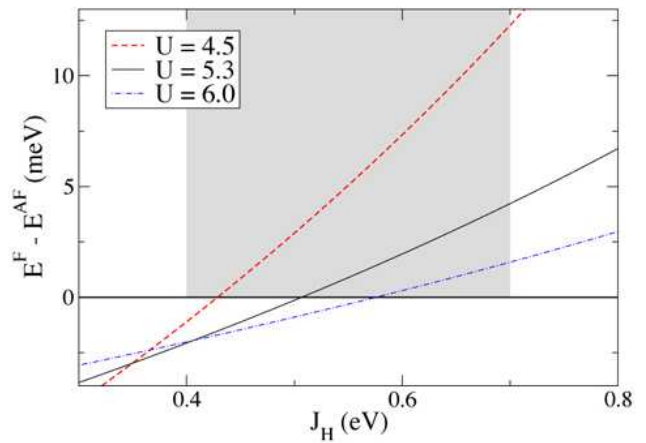}}
\caption{Energy gain $E^{\rm F} - E^{\rm AF}$ of a nearest
neighbor Co2 ferromagnetic alignment
with respect to an antiferromagnetic orientation as a function of the
Hund's coupling $J_{\rm H}$, for a typical parameter set of
Hubbard $U$. The grey shaded area indicates the interval of
$J_{\rm H}$ for which a Co2-high-spin Co1-low-spin configuration can
be stabilized for the considered crystal field
splittings.\cite{Laschinger03} }
\label{Fig:stability}
\end{figure}

We proceed with a strong coupling expansion to identify the magnetic
coupling along the chain. This amounts to determine
the difference in energy, between 
the ferromagnetic and antiferromagnetic configurations, to fourth
order in the hopping, since this is the leading order to the magnetic
interaction between the high spin Co ions. As explained above we only
have to take into account the $3z^2$-$r^2$ level on Co2 and the
$t_{2g}$ levels on Co1. In an ionic picture all $t_{2g}$ levels on Co1
are filled while the $d_{3z^2-r^2}$ level on Co2 is half-filled and we
therefore consider hopping processes from the former to the latter. In the
ferromagnetic configuration we include processes where two down spin
electrons hop from Co1 to both neighboring Co2 and back
again as displayed 
in Fig.~\ref{Fig:levelscheme}a. There are in total 
$3\times 2\times 2\times 2\times 2 = 48$ such processes. The
intermediate spin state for Co1 is in agreement with 
Hund's rule. The energy gain per path is given by: 
\begin{equation}
{ E_f=\frac{t^4}{E_0^2} \;
\left(3U-5J_{\rm{H}}+4 E_{\rm loc}(\Delta_{\rm{cf}},J_{\rm{H}},-1)\right)^{-1}}
\end{equation}
with
\begin{equation}
{ E_0 = U-J_{\rm{H}}+2E_{\rm loc}(\Delta_{\rm{cf}},J_{\rm{H}},-2)}
\end{equation}
\begin{equation}
\Delta_{\rm{cf}} = \Delta_{\rm{Co1}} + \frac{4}{10} \Delta_{\rm{Co2}}
\end{equation}
and
\begin{equation}
{ E_{\rm loc}(\Delta_{\rm{cf}},J_{\rm{H}},l)}=
   {\frac{\Delta_{\rm{cf}}J_{\rm{H}}^2}{(\Delta_{\rm{cf}}-
   \frac{1}{2}lJ_{\rm{H}})(2\Delta_{\rm{cf}}+3J_{\rm{H}})}}
\end{equation}
where $\Delta_{\rm{Co1}}$ and $\Delta_{\rm{Co2}}$ denote the
crystal field splittings on Co1 and Co2, respectively. 
The Hund's coupling is $J_{\rm{H}}$, assumed to be identical on both,
Co1 and Co2, and $U$ denotes the local Coulomb repulsion. 
There are no further paths in this configuration, besides
the one which twice iterates second order processes. In the antiferromagnetic
case the situation is slightly more involved. Here three different
classes of paths have to be distinguished. The first class, denoted $a1$ in the
following, consists of hopping events of one up spin and
one down spin electron from the same Co1 level. (There are $3\times
2\times 2 = 12$ 
such paths). The second class ($a2$) consists of hopping events of one
down spin and 
one up spin electron from different Co1 levels (There are $3\times
2\times 2 \times 2  = 24$ such paths). The third class ($a3$), shown
in Fig.~\ref{Fig:levelscheme}b, consists
of hopping processes where one electron is hopping from Co1 to Co2 and
then  another electron is hopping from the other Co2 to the same Co1
and back again (There are $3\times 2= 6 $ such paths). In total this
sums up to 42 paths in the antiferromagnetic 
configuration. Consequently, we have more ferromagnetic than
antiferromagnetic exchange paths.
However the energy gain depends on the 
path. For the classes $a1$ and $a2$, the intermediate Co1 state 
violates the Hund's rule,
and we identify an energy gain per path given by:
\begin{eqnarray}\nonumber
E_{a1}&=&\frac{t^4}{E_0^2}\left(3U-5J_{\rm{H}}+(4-\frac{6J_{\rm{H}}}{\Delta_{\rm{cf}}})
E_{\rm loc}(\Delta_{\rm{cf}},J_{\rm{H}},1)\right)^{-1}\\
E_{a2}&=&\frac{t^4}{E_0^2}\left(3U-2J_{\rm{H}}-\rm{F}(\Delta_{\rm{cf}},J_{\rm{H}})\right)^{-1}
\end{eqnarray}
Here F is a positive function which is smaller than
$J_{\rm{H}}$. The expression $3U-2J_{\rm{H}}-F$ is the 
lowest eigenvalue of $\langle i|H_{\rm Co}|j \rangle$ where the states $i$ and
$j$ are all possible states on  Co1 consistent with two of the $d$-orbitals
filled and three empty.
For the class $a3$ one observes that one does not need to invoke a Co1
ion with four electrons as an intermediate state, in contrast to
all other processes we considered so far. We find the energy
gain as: 
\begin{equation}
E_{a3}=\frac{t^4}{E_0^2\left(U+2J_{\rm H}\right)}\\
\end{equation}
Altogether we obtain the difference in energy gain between the
ferromagnetic and the antiferromagnetic configurations as:
\begin{equation}
E^{\rm F} - E^{\rm AF}=48 E_f-24E_{a1}-12E_{a2}-6E_{a3}
\end{equation}
The dependence of $E^{\rm F} - E^{\rm AF}$  on $J_{\rm H}$ for different values
of the local interaction $U$ is shown in Fig.~\ref{Fig:stability}.
Using $J_{\rm H}=0.6~\rm{eV}$,\cite{Laschinger03}
$U=5.3~\rm{eV}$,\cite{Sawatzky91}  $t=1.5~\rm{eV}$,\cite{Eyert03}
$\Delta_{\rm{Co1}}=2.5~\rm{eV}$ and $\Delta_{\rm{Co2}}=1.5~\rm{eV}$,\cite{Eyert03}
we obtain an estimate for the Heisenberg exchange coupling
(for the Co2 spin $S=2$):
\begin{equation}
J^{\rm F}=(E^{\rm F} - E^{\rm AF})/2S^2\approx 2 \rm{meV} 
\end{equation}
which is in reasonable agreement with the experimental transition
temperature of 25~K. 

In this context one should realize that a
one-dimensional chain does not support a true phase transition into
the magnetic state. However, as the length  $L$ of the chains is finite,
a crossover into the ferromagnetic state may be observed
when the correlation length is approximately $L$.\cite{chain}

To emphasize 
the importance of the chain geometry
we now briefly discuss the
hypothetical case where the $z$-axis of the octahedra corresponds to 
that of the prism. In this geometry there is only one orbital on 
each Co ion 
which contributes to the hopping processes. In this situation the
process favoring ferromagnetism shown in Fig.~\ref{Fig:levelscheme}a does not
exist, in contrast to the process $a3$ shown in
Fig.~\ref{Fig:levelscheme}b, and the resulting coupling is therefore
antiferromagnetic.  

In the large class of known isostructural
compounds\cite{Stitzer01} the non-magnetic ion is not necessarily a Co
ion. If the non-magnetic ion is in a
$3d^2$  (or $4d^2$)  configuration, the above argument applies, 
and the coupling is  antiferromagnetic. 
If the configuration is $3d^4$, all the discussed electronic processes
contribute, however with different multiplicities. 
Moreover, additional paths have to be considered for the
antiferromagnetic case. They represent exchange processes through an empty
orbital on the non-magnetic ion.
As a result, the ferromagnetic scenario has fewer paths than the
antiferromagnetic, and the coupling becomes antiferromagnetic.
Correspondingly, a ferromagnetic interaction can only occur when all
three orbitals on the nonmagnetic ion
participate in the exchange process. Obviously the situation 
we consider differs from the standard
180~degree superexchange mechanism in many respects.   

With the investigation of the {\it interchain} magnetic interaction one first notices
that each magnetic Co ion has twelve neighboring Co ions on different
chains. However, as displayed in Fig.~\ref{Fig:plane}, there is an
oxygen bridge to  only six neighbors, one per chain. Here the coupling 
$J^{\rm AF}$ results from the super-superexchange
mechanism (with exchange via two oxygen sites), and it is
antiferromagnetic. Since the Co-O hybridization is unusally large 
in this system, we expect the interchain magnetic coupling to be 
sufficiently strong to account for the observed antiferromagnetic 
correlations.

From our previous considerations we introduce the minimal magnetic Hamiltonian:
\begin{eqnarray}\label{Eq:Hamiltonian}  
H &=& \sum_{i,j} \left(J^{\rm F}_{i,j} {\vec S}_i \cdot {\vec S}_j + J^{\rm AF}_{i,j}
{\vec S}_i \cdot {\vec S}_j \right) -D \sum_{i} S_{z,i}^2 \\
{\rm with}&& J^{\rm F}_{i,j} = \left\{ \begin{array}{ll}
J^{\rm F}& \mbox{if ${\vec j} - \vec{i} = \pm 2 {\vec d}$}\\
0& \mbox{otherwise}
\end{array} \right. \nonumber \\
{\rm and} \nonumber\\
J^{\rm AF}_{i,j}\! &=& \!\! \left\{ \begin{array}{ll}
J^{\rm AF}& \mbox{if ${\vec j} - \vec{i} = \pm ( {\vec a}+{\vec d}), \pm ( {\vec
b}+{\vec d}), \pm ( {\vec c}+{\vec d}) $}\nonumber \\
0& \mbox{otherwise.}
\end{array} \right. \\ \nonumber
\end{eqnarray}
Here we use the site vectors ${\vec a} = a (-1/2,\sqrt{3}/2,c/(12 a))$, ${\vec b} = a
(-1/2,-\sqrt{3}/2,c/(12 a))$, ${\vec c} = a (1,0,c/(12 a))$, and
${\vec d} = {\vec a}+{\vec b}+{\vec c}$ where $a=9.06$~\AA\ and 
$c=10.37$~\AA\ are the lattice constants of the hexagonal unit
cell. The Hamiltonian, Eq.~(\ref{Eq:Hamiltonian}), also includes a
phenomenological contribution $D S_{z}^2$
which accounts for the anisotropy  observed, for example, in the
magnetic susceptibility.\cite{Kageyama97a,Maignan00}

The stability of the magnetization steps results from the magnetic
frustration which is introduced through the antiferromagnetic
interchain coupling. Indeed, the lattice structure suggests that this
magnetic system is highly frustrated, since the chains are arranged on a
triangular lattice. However investigating the
Hamiltonian~(\ref{Eq:Hamiltonian}) 
reveals that the microscopic mechanism leading
to frustration is more complex. It is visualized when we
consider a closed path $TLRT'T$ --- where the sites $T$ and $T'$ are
next nearest-neighbor Co2 sites on the same chain and the sequence of 
sites $TLR$ is located on a triangle of
nearest neighbor chains. One advances from $T$ to $L$ to $R$ to $T'$
on a helical path\cite{comment} formed by the oxygen bridges from
Fig.~\ref{Fig:plane}. Since the structure imposes $T$ and $T'$ to be
next nearest-neighbors, the frustration occurs independently of the
sign of the intrachain coupling.

In summary we established the magnetic interactions in 
an effective magnetic Hamiltonian for
Ca$_3$Co$_2$O$_6$. It is a spin-2 Hamiltonian, with antiferromagnetic
interchain coupling, and ferromagnetic intrachain interactions. The
latter is obtained from the evaluation of all spin exchange paths
between two high-spin Co2 sites through an intermediary low-spin
Co1 site.  This mechanism is particular to the geometry of the
system as is the microscopic mechanism which leads to magnetic
frustration. 
We expect that the discussed microscopic mechanisms also 
apply to other isostructural compounds, such as
Ca$_3$CoRhO$_6$ and Ca$_3$CoIrO$_6$.

\begin{acknowledgments}
We are grateful to A.~Maignan, C.~Martin, Ch.~Simon, C.~Michel,
A.~Guesdon, S.~Boudin and V.~Hardy for useful discussions.
C.~Laschinger is supported by a Marie Curie fellowship of the European
Community program under number HPMT2000-141.  The
project is supported by DFG through SFB~484 and by 
BMBF~(13N6918A).
\end{acknowledgments}


\begin{references}
\bibitem{chain}
Given the strong magnetic anisotropy of
Ca$_3$Co$_2$O$_6$,\cite{Maignan00} 
which is phenomenologically included in Eq.~(\ref{Eq:Hamiltonian}), 
we expect that one can capture some physics of the system in the Ising
limit. 
In that case, for a typical chain length 
of the order of one hundred sites, the correlation length 
extends to the system size for $T\simeq 0.8 J$ which is to be
interpreted as the crossover temperature to the ferromagnetic state.
Correspondingly, the spin susceptibility peaks at about
$T/J\simeq0.8-1.2$ which we verified in exact diagonalization. 
Even with chain lengths of 10,000 Co sites, the crossover temperature 
is still in the given range.
\bibitem{comment}
 	The positions of the considered sites are:
               $T$ is an arbitrary Co2 site on any chain,
               $L=T+{\vec b}+{\vec d}$, $R=L+{\vec c}+{\vec d}$ 
               and $T'=R+{\vec a}+{\vec d}$.
\end{references}

\end{document}